\documentclass{elsart}

\usepackage{amssymb,amsmath}

\begin{document}

\newcommand{\vspacebefore}{\raisebox{0ex}[2.5ex][0ex]{\null}}
\newcommand{\vspacebeforemore}{\raisebox{0ex}[3.5ex][0ex]{\null}}
\newcommand{\p}{\partial}
\newcommand{\const}{\mathop{\rm const}\nolimits}
\newcommand{\Equiv}{\mathop{\rm \, equiv}}
\newcommand{\sign}{\mathop{\rm sign}\nolimits}

\def\X#1{X_{\rm #1}}

\renewcommand{\i}{{\rm i}}

\newcounter{tbn}
\newcounter{tabul}

\journal{JMAA}

\begin{frontmatter}

\title{Conservation Laws and Symmetries\\ of Semilinear Radial Wave Equations}

\author{Stephen C. Anco}
\ead{sanco@brocku.ca}
\address{Department of Mathematics, Brock University, St. Catharines, ON, L2S 3A1, Canada}

\author{Nataliya M. Ivanova}
\ead{ivanova@imath.kiev.ua}
%ivanova@math.ubc.ca
\address{Institute of Mathematics of NAS of Ukraine, 01601 Kyiv, Ukraine\\
and\\
Department of Mathematics, University of British Columbia, Vancouver, BC,\\ V6T 1Z2, Canada}

\begin{abstract}
Classifications of symmetries and conservation laws are presented for a 
variety of physically and analytically interesting wave equations 
with power nonlinearities
in $n$ spatial dimensions:
a radial hyperbolic equation,
a radial Schr\"odinger equation and its derivative variant,
and two proposed radial generalizations of modified
Korteweg--de Vries equations,
as well as Hamiltonian variants.
The mains results classify all admitted local point symmetries
and all admitted local conserved densities depending on
up to first order spatial derivatives,
including any that exist only for special powers or dimensions.
All such cases for which these wave equations admit, in particular,
dilational energies or conformal energies and inversion symmetries
are determined.
In addition, potential systems arising from the classified conservation laws
are used to determine nonlocal symmetries and nonlocal conserved quantities
admitted by these equations.
As illustrative applications, a discussion is given of
energy norms, conserved $H^s$ norms,
critical powers for blow-up solutions,
and one-dimensional optimal symmetry groups for invariant solutions.
\end{abstract}


\end{frontmatter}

\section{Introduction}

Over past few decades there has been a lot of work on global analysis of
nonlinear wave equations
in $n\ge1$ spatial dimensions \cite{Strauss1989},
\[
u_{tt}=\Delta u+f(x,u,\nabla u) ,\quad
u(t,x) \in {\mathbb R} , \tag{WEa}
\]\vspace{-4ex}
\[
\i u_{t}=\Delta u+f(x,|u|,|\nabla u|)u ,\quad
u(t,x) \in {\mathbb C} ,  \tag{WEb}
\]
with one main focus being the study of blow-up phenomenon
for the case of power nonlinearities $f=\pm |u|^p$, $\pm|\nabla u|^p$.
Some key tools with a number of uses in this study are
conservation laws and symmetries.

Conservation laws such as energy provide basic conserved quantities
used in obtaining estimates on $|u|$ or $|\nabla u|$
for smooth solutions,
and also in defining suitable norms for weak solutions.
Of considerable interest are extra conservation laws
such as conformal energies that can appear for special powers $p$
depending on the dimension $n$.
Symmetries, in contrast, lead to exact group-invariant solutions
and play a role in defining invariant Sobolev norms.
Scaling symmetries are of special relevance,
as the critical nonlinearity power for blow-up is typically singled out by
scaling-invariance of a positive energy norm.
Moreover, scaling transformation arguments give a means of
relating the behavior of solutions
in different regimes, for instance,
solutions at short times with large initial data
can be scaled to long times with small initial data
when the nonlinearity power is subcritical.

In this paper
we present classifications of conservation laws and symmetries
admitted by a variety of physically and analytically interesting
semilinear radial wave equations.
The equations will be organized according to their variational structure
as follows.

First, we consider the standard (hyperbolic) nonlinear wave equation
\[
u_{tt}=\Delta u\pm u^p
\tag{NLW}
\]
as well as the nonlinear Schr\"odinger equation
\[
\i u_t=\Delta u\pm |u|^p u
\tag{NLS}
\]
and its radial derivative variant
\[
\i u_t=\Delta u\pm \i |u|^p\Big(u_r+\frac m{p+2} r^{-1}u \Big)
\tag{dNLS}
\]
where $\Delta=r^{-m}\p_r r^m \p_r =\p_r^2+m r^{-1}\p_r$
is the radial Laplacian in $n=m+1\ge1$ (spatial) dimensions.
These PDEs each arise as the stationary points
$\delta\mathcal{L}/\delta u=0$ of a Lagrangian functional
$\mathcal{L}=\int_{-\infty}^{+\infty}\int_0^{\infty} L[u] r^m dr dt$,
given by
\begin{align*}
L_{\rm NLW}[u]&=\frac{1}{2}(-u_t^2+u_r^2)\mp\frac{1}{p+1}u^{p+1},\\
L_{\rm NLS}[u]&=\i\bar uu_t + |u_r|^2 \mp\frac{1}{p+2}|u|^{p+2},\\
L_{\rm dNLS}[u]&=\i\bar uu_t + |u_r|^2\mp\frac{\i}{p+2}|u|^p
(\bar uu_r -u\bar u_r).
\end{align*}

As is well known, the NLS equation also possesses a Hamiltonian formulation
$u_t=\i\delta \mathcal{H}/\delta\bar u$
with
$\mathcal{H}=\int_0^\infty H[u] r^mdr$ given by
\[
H_{\rm NLS}[u]=-|u_r|^2\pm\frac{2}{p+2} |u|^{p+2}
\]
where multiplication by $\i$
defines a Hamiltonian operator
with respect to the $L^2$ Hermitian inner product on the radial line
$0\leq r<\infty$.
Next, while the derivative version (dNLS) of the NLS equation
is not Hamiltonian,
it does have a Hamiltonian variant
\[
\i u_t=
\Delta u+\frac{m(m-2)}4 r^{-2}u
\pm \i\left((|u|^pu)_r+\frac{m}{2} |u|^p r^{-1}u \right)
\tag{dNLS-H}
\]
which reduces to the standard derivative NLS equation in the case $m=0$.
This radial generalization
arises from the Hamiltonian formulation
\begin{equation}\label{HamFormulation}\textstyle
u_t=r^{-m/2}\p_r(r^{-m/2}\delta \mathcal{H}/\delta\bar u),\quad
\mathcal{H}=\int_0^\infty H[u] r^mdr
\end{equation}
with
\[
H_{\rm dNLS}[u]=\frac{\i}{2}(u\bar u_r-\bar u u_r) \pm\frac{2}{p+2}|u|^{p+2}.
\]
Here $r^{-m/2}\p_rr^{-m/2}$ is easily verified to be a Hamiltonian operator
with respect to the radial $L^2$ Hermitian inner product.
In particular, 
such an operator $\mathcal{D}$ 
with no dependence on $u$ or $\bar u$ or their derivatives
is {\em Hamiltonian} \cite{Olver1986}
iff it is skew-adjoint in this inner product
so consequently the Poisson bracket associated to it by
$\{\mathcal{P},\mathcal{Q}\}_\mathcal{D}
= \int_0^\infty (\delta \mathcal{P}/\delta \bar u)
\mathcal{D}(\delta \mathcal{Q}/\delta u) dr$
will be skew-Hermitian
%\[
%\overline{\{\mathcal{P},\mathcal{Q}\}}_\mathcal{D}= - \{\mathcal{Q},\mathcal{P}\}_\mathcal{D}
%\]
and obey the Jacobi identity
%\[
%\{\{\mathcal{P},\mathcal{Q}\}_\mathcal{D},\mathcal{R}\}_\mathcal{D}+\rm{cyclic} =0,
%\]
for all real functionals
$\mathcal{P}=\int_0^\infty P[u] r^m dr$,
$\mathcal{Q}=\int_0^\infty Q[u] r^m dr$.
Note that, as a consequence of the skew-adjoint property, 
the Hamiltonian $\mathcal{H}=\int_0^\infty H[u] r^mdr$
will formally be a conserved quantity,
$\frac{d}{dt}\mathcal{H}=0$
(to within boundary terms at spatial infinity),
for all formal solutions $u$.

Last, based on the factorization of the Laplacian
$\Delta=(\p_r+m r^{-1})\p_r$,
we propose two radial generalizations of
the modified Korteweg--de Vries equation,
\[
u_t=(\Delta u\pm u^{p+1})_r
\tag{mKdV-1}
\]
and
\[
u_t=(\Delta u\pm u^{p+1})_r +\frac{m}{r}(\Delta u\pm u^{p+1})
\tag{mKdV-2}
\]
both of which have neither a Lagrangian nor Hamiltonian formulation
except in the case $m=0$.
We also introduce a Hamiltonian variant given by
\[
u_t=(\Delta u\pm u^{p+1})_r +\frac{m}{2r}(\Delta u\pm u^{p+1})
\tag{mKdV-H}
\]
with
\[
H_{\rm mKdV}[u] = -\frac{1}{2} u_r^2 \pm \frac{1}{p+2} u^{p+2}
\]
using the previous Hamiltonian operator~\eqref{HamFormulation},
specialized in the obvious way to real functions $u=\bar u$.
These radial mKdV equations are examples of third order evolutionary
wave equations
\[
u_{t}=\Delta \nabla_x u+f(x,u,\nabla u,\Delta u) ,\quad
u(t,x) \in {\mathbb R} , \tag{WEc}
\]
with power nonlinearities $f=u^p \nabla_x u$, $u^{p+1}$,
where $\nabla_x =|x|^{-1}x\cdot\nabla$ is the radial gradient.
Well-posedness of such wave equations is an interesting problem
which has received some recent attention.

We emphasize the wave equations (dNLS), (dNLS-H), (mKdV-1,2,H)
for $m\neq 0$ are new, being radial generalizations of the familiar
($n=1$ dimensional) derivative Schr\"odinger equation
and modified Korteweg--de Vries equation.

To begin we recall the definitions of symmetries and conservation laws
from an analytical perspective
(see also \cite{Olver1986,Bluman&Anco2002}).
A {\em point symmetry} of a radial wave equation~(WE)
is a group of transformations given by an infinitesimal generator
\[
\delta t=\tau(t,r,u),\qquad \delta r=\xi(t,r,u), \qquad \delta u=\eta(t,r,u)
\]
on the variables $t$, $r$, $u$
in the real-valued case,
or
%\begin{gather*}
\[
\delta t=\tau(t,r,u,\bar u),\quad \delta r=\xi(t,r,u,\bar u), \quad
%\\
\delta u=\eta(t,r,u,\bar u), \quad \delta\bar u=\bar\eta(t,r,u,\bar u)
\]
%\end{gather*}
on the variables $t$, $r$, $u$ and $\bar u$
in the complex-valued case,
such that the equation~(WE) is preserved.
On solutions, such a transformation in both cases
is infinitesimally equivalent to
\begin{equation}\label{pointsymm}
\delta t=\delta r=0, \qquad \delta u=\eta-\tau u_t-\xi u_r
\end{equation}
called the {\em characteristic form} of the point symmetry.
The expressions $\eta$, $\tau$, $\xi$ are determined by
the Fr\'echet derivative of the wave equation~(WE)
applied to $\delta u$ holding for all formal solutions $u$.
(More precisely,
one works in a jet space setting,
using the coordinate space defined by
$t$, $r$, $u$ and all derivatives of $u$
modulo the equation~(WE) and its differential consequences.
In jet space, a point symmetry is the prolongation of the operator
$X =\tau\partial_t+\xi\partial_r+\eta\partial_u$.
When $u$ is complex-valued, jet space is enlarged in the obvious way
by $\bar u$ and its derivatives.)
The set of all infinitesimal point symmetries admitted by
a given wave equation~(WE) has the structure of a Lie algebra
(for the operators $X$ under commutation).
For a given point symmetry,
invariant solutions are characterized by the form
$\delta u=\eta(t,r,u)-\tau(t,r,u)u_t-\xi(t,r,u)u_r=0$
when $u(t,r) \in {\mathbb R}$, or
$\delta u=\eta(t,r,u,\bar u)-\tau(t,r,u,\bar u)u_t-\xi(t,r,u,\bar u)u_r=0$
together with its complex conjugate
when $u(t,r) \in {\mathbb C}$.

A {\em conservation law} of a radial wave equation~(WE)
is given by a space-time divergence $D_t(r^m\Psi^t)+D_r(r^m\Psi^r)$
that is equal to a linear combination of
the equation and its differential consequences,
so that
\begin{equation}\label{conslaw}\textstyle
D_t\Psi^t+D_r\Psi^r +mr^{-1}\Psi^r=0
\end{equation}
holds for all formal solutions~$u$.
The radial integral of the conserved density~$\Psi^t$ formally satisfies
\[
\frac d{dt} \int_0^\infty\Psi^t r^m dr =-r^m\Psi^r|_0^\infty =0
\]
which vanishes
when the flux~$\Psi^r$ at spatial infinity is zero
or decays faster than $r^{-m}$.
Hence $C=\int_0^\infty\Psi^t r^m dr$
formally yields a conserved quantity for the equation~(WE).
Conversely, any such conserved quantity arises from
a conservation law~\eqref{conslaw}.
Two conservation laws are equivalent if their conserved densities~$\Psi^t$
differ by a radial divergence $r^{-m}D_r(r^m \Theta)$
on all formal solutions~$u$,
giving the same conserved quantity $C$ up to boundary terms.
The set of all conservation laws (up to equivalence)
admitted by a given wave equation~(WE)
forms a vector space,
on which there is a natural action \cite{Bluman&etal}
by the Lie group of
all admitted point symmetries of the equation~(WE).

Each conservation law~\eqref{conslaw} of a radial wave equation~(WE)
has an equivalent {\em characteristic form}
where $D_t(r^m\Psi^t)+D_r(r^m\Psi^r)$ is just
proportional to the equation (WE)
multiplied by an expression $Q$ that depends on the jet variables
(up to some finite differential order).
Such expressions $Q$ for which the product of equation (WE) and $Q$ yields
a total space-time divergence 
(and hence a conservation law on solutions of the equation (WE)) 
are called {\it multipliers}. 
There is a specific relation between multipliers $Q$
and conserved densities $\Psi^t$:
in the case of a hyperbolic wave equation~(WEa),
conserved densities $\Psi^t(t,r,u,u_r,u_t)$
modulo radial divergences
correspond to multipliers
\begin{equation}\label{multiplierWEa}\textstyle
Q(t,r,u,u_r,u_t) =\delta(\Psi^t r^m)/\delta u_t ,
\end{equation}
while in the case of an evolutionary wave equation~(WEb) and (WEc),
conserved densities $\Psi^t(t,r,u,u_r)$
or $\Psi^t(t,r,u,\bar u,u_r,\bar u_r)$
modulo radial divergences
correspond to multipliers
\begin{gather}
Q(t,r,u,u_r,u_{rr})=\delta(\Psi^t r^m)/\delta u ,
\label{multiplierWEb}\\ 
Q(t,r,u,\bar u,u_r,{\bar u}_r,u_{rr},{\bar u}_{rr})
=-\i \delta(\Psi^t r^m)/\delta u ,
\label{multiplierWEc}
\end{gather}
respectively.
In all cases the multiplier $Q$ is determined by
\cite{Anco&Bluman2002a,Anco&Bluman2002b}
the adjoint of the Fr\'echet derivative of equation~(WE)
applied to $Q$,
augmented by additional equations formed from
the Fr\'echet derivative of $Q$ itself,
holding for all formal solutions~$u$.
(More precisely,
one works in the same jet space as for the computation of symmetries.)
Thus the determination of conservation laws via multipliers is
a kind of adjoint problem \cite{Anco&Bluman1997}
of the determination of symmetries.

If a wave equation~(WE) possesses a Lagrangian formulation, 
then its multipliers $Q$ define {\em variational} symmetries 
$\delta t=\delta r=0$, $\delta u=Q$ in the real-valued case
or
$\delta t=\delta r=0$, $\delta u=\bar Q$ in the complex-valued case,
such that the determining equations on $Q$
reduce to conditions equivalent to those given by Noether's theorem
\cite{Olver1986,Anco&Bluman2002a,Anco&Bluman2002b}
for the Lagrangian to be formally invariant
(to within boundary terms at spatial infinity).
Variational symmetries corresponding to multipliers
\eqref{multiplierWEa}, \eqref{multiplierWEb}, \eqref{multiplierWEc}
for conserved densities containing derivatives of $u$ or $\bar u$
will be a point symmetry $\delta u=Q$ in the real-valued case
only if the derivatives of $u$ in $Q$ at most
are first order and appear linearly,
or in the complex-valued case $\delta u= \bar Q$ will be a point symmetry
only if the same is true for derivatives of $\bar u$ in $Q$
and there are no derivatives of $u$ in $Q$.

For each of the wave equations
(NLW), (NLS), (dNLS) and (dNLS-H), (mKdV-1,2,H),
we now classify all admitted conserved quantities
containing up to first order spatial derivatives,
along with all admitted point symmetries.
In these classifications,
the nonlinearity power will be restricted to
$p\neq 0,1$ for (NLW),
and $p\neq 0$ for (NLS), (dNLS), (dNLS-H), (mKdV-1,2,H),
so that all cases of linear wave equations are excluded.
No restrictions will be placed on $m$
(even allowing non-integer values).
All computations have been carried out using the computer algebra programs
{\sc LiePDE} and {\sc ConLaw} \cite{Wolf1993,Wolf2002b}
utilizing an enhanced version of the program {\sc Crack} \cite{Wolf2002a}
for solving overdetermined systems of equations containing parameters.

Throughout we use the notation
%$F(u, q)=\frac1q u^q,$ for $q\ne 0$; $\ln u,$ for $q=0$.
$F(u, q)=
%\begin{cases}
\big\{\begin{smallmatrix}
\frac1q u^q,\ q\ne 0\\
\ln u,\ q=0
\end{smallmatrix}
%\end{cases}
$.
Also, a ``$\pm$'' sign will refer to the sign of the nonlinear term
in the wave equations, usually called the {\em focusing/defocusing} cases,
respectively.

\section{ Point symmetries }

All of the wave equations
(NLW), (NLS), (dNLS) and (dNLS-H), (mKdV-1,2,H)
obviously admit time translations $\tau=1$, $\xi=\eta=0$,
and, for $m=0$, space translations $\xi=1$, $\tau=\eta=0$.
In addition the Schr\"odinger equations (NLS), (dNLS), (dNLS-H)
admit phase rotations $\tau=\xi=0$, $\eta=\i u$.

The following tables list,
firstly, the scaling symmetries admitted by these equations,
and secondly, any extra admitted point symmetries.
(Note these classifications exclude all linear cases i.e.
$p=0$; plus $p=1$ for (NLW).)

\newpage

\setcounter{tbn}{0}\setcounter{tabul}{0}
{\begin{center}\refstepcounter{tabul}\label{TableScaleSym}
Table~\thetabul. Scaling symmetries
\\[1.5ex] \scriptsize
\setcounter{tbn}{0}
\begin{tabular}{|l|l|l|l|}
\hline\vspacebefore\scriptsize
\hfill $\tau\hfill$ &\hfill $\xi\hfill$ &\hfill$\eta\hfill$ &\hfill {Equation\hfill} \\
\hline\vspacebeforemore
 $t$ & $r$ &$-\dfrac2{p-1}u$ & (NLW)  \\[1.2ex]
\hline\vspacebeforemore
 $2t$ & $r$ &$-\dfrac2p u$  & (NLS)\\[1.2ex]
\hline\vspacebeforemore
$2t$ & $r$ &$-\dfrac1{p} u$  &  (dNLS),  (dNLS-H)\\[1.2ex]
\hline\vspacebeforemore
$3t$ & $r$ &$-\dfrac2pu$  &  (mKdV-1,2,H)\\[1.2ex]
\hline
\end{tabular}
\end{center}}

%\newpage

{\begin{center}\refstepcounter{tabul}\label{TablePointSymNLW}
Table~\thetabul. Extra point symmetries for (NLW)
\\[1.5ex] \scriptsize
\setcounter{tbn}{0}
\begin{tabular}{|l|l|l|l|l|}
\hline\vspacebefore
\hfill $\tau\hfill$ &\hfill $\xi\hfill$ &\hfill$\eta\hfill$ &\hfill {Remarks\hfill} \\
\hline\vspacebefore
$r$ & $t$ & 0 & Lorentz boost, $m=0$\\
\hline
{\raisebox{0ex}[3.5ex][0ex]{\null}}$t^2+r^2$ & $2tr$ & $-\dfrac{4}{p-1}tu$ &
    inversion, $m=\dfrac4{p-1}$\\[1.2ex]
\hline
\end{tabular}
\end{center}}

\setcounter{tbn}{0}
{\begin{center}\refstepcounter{tabul}\label{TablePointSymNLS}
Table~\thetabul. Extra point symmetries for (NLS)
\\[1.5ex] \scriptsize
\setcounter{tbn}{0}
\begin{tabular}{|l|l|l|l|l|}
\hline\vspacebefore
\hfill $\tau\hfill$ &\hfill $\xi\hfill$ &\hfill$\eta\hfill$   &\hfill {Remarks\hfill} \\
\hline\vspacebefore
 0 & $2t$ &$\i ru$ & Galilean boost, $m=0$\\
\hline\vspacebeforemore
 $t^2$ & $tr$ &$\Big(\dfrac{\i}{4} r^2-\dfrac2p t\Big)u$ & inversion, $m=-1+\dfrac4p$\\[1.2ex]
\hline
\end{tabular}
\end{center}}

The inversion and boost symmetries of equations~(NLW) and (NLS)
are well known~\cite{Strauss1989},
so our classification in tables~\ref{TablePointSymNLW} and~\ref{TablePointSymNLS}
mainly provides a completeness result
that these two equations do not possess any additional symmetries
for special nonlinearity powers.

Surprisingly, no extra symmetries are found to be admitted by
equations~(dNLS) and (dNLS-H).

\setcounter{tbn}{0}
{\begin{center}\refstepcounter{tabul}\label{TablePointSymmKdV-1}
Table~\thetabul. Extra point symmetries for (mKdV-1)
\\[1.5ex] \scriptsize
\setcounter{tbn}{0}
\begin{tabular}{|l|l|l|l|}
\hline\vspacebefore
\hfill $\tau\hfill$ &\hfill $\xi\hfill$ &\hfill$\eta\hfill$ &\hfill {Remarks\hfill} \\
\hline\vspacebefore
 $0$ & $2t$ & $\mp1$& Galilean boost, $m=0$, $p=1$ \\[0.5ex]
\hline
\end{tabular}
\end{center}}

%\newpage

\setcounter{tbn}{0}
{\begin{center}\refstepcounter{tabul}\label{TablePointSymmKdV-2}
Table~\thetabul. Extra point symmetries for (mKdV-2)
\\[1.5ex] \scriptsize
\setcounter{tbn}{0}
\begin{tabular}{|l|l|l|l|}
\hline\vspacebefore
\hfill $\tau\hfill$ &\hfill $\xi\hfill$ &\hfill$\eta\hfill$ &\hfill {Remarks\hfill} \\
\hline\vspacebefore
$0$ & $2t$ & $\mp1$& Galilean boost, $m=0$, $p=1$\\
\hline\vspacebeforemore
 $t^2$ & $\dfrac23tr$ & $-\dfrac13(4tu\pm r)$& inversion, $m=1$, $p=1$\\[1.2ex]
\hline
\end{tabular}
\end{center}}

\setcounter{tbn}{0}
{\begin{center}\refstepcounter{tabul}\label{TablePointSymmKdV-H}
Table~\thetabul. Extra point symmetries for (mKdV-H)
\\[1.5ex] \scriptsize
\setcounter{tbn}{0}
\begin{tabular}{|l|l|l|l|}
\hline\vspacebefore
\hfill $\tau\hfill$ &\hfill $\xi\hfill$ &\hfill$\eta\hfill$ &\hfill {Remarks\hfill} \\
\hline\vspacebefore
 0 & $2t$ & $\mp1 $ & Galilean boost, $m=0$, $p=1$ \\[0.5ex]
\hline\vspacebeforemore
 $t^2$ & $\dfrac23tr$ &$-\dfrac13(4tu\pm r)$ & inversion, $m=2$, $p=1$\\[1.2ex]
\hline
\end{tabular}
\end{center}}

Like the (NLS) equation, the mKdV equations~(mKdV-2,H) possess
boosts and inversions as extra symmetries for special nonlinearity powers,
while no extra symmetries are found to be admitted by equation (mKdV-1).
Our results in tables~\ref{TablePointSymmKdV-1},~\ref{TablePointSymmKdV-2},~\ref{TablePointSymmKdV-H}
for these radial mKdV equations with $m>0$ are new.

Note the powers for which the inversions exist are called a {\em conformal power}.

%\newpage

\setcounter{tbn}{0}
{\begin{center}\refstepcounter{tabul}\label{TableConfPowers}
Table~\thetabul. Conformal powers
\\[1.5ex] \scriptsize
\setcounter{tbn}{0}
\begin{tabular}{|l|l|}
\hline\vspacebefore
\hfill $p\hfill$  &\hfill {Equation\hfill} \\
\hline\vspacebeforemore
 $1+\dfrac4m$ & (NLW), $m\ne0$  \\[1.2ex]
\hline\vspacebeforemore
  $\dfrac4{m+1}$ & (NLS), $m\ne-1$ \\[1.2ex]
\hline\vspacebefore
1 & (mKdV-2), $m=1$; (mKdV-H), $m=2$\\
\hline
\end{tabular}
\end{center}}

As a summary of the main results,
we list the point symmetry algebras found for $m>0$,
with their generators denoted by 
$\X{trans}$, $\X{scal}$, $\X{inver}$, $\X{phase}$.
These algebras fall into four classes
(with a suitable normalization of $\X{scal}$ in each case):
For equation (mKdV-1)
as well as the non-conformal case of equations (NLW) and (mKdV-2,H),
the admitted algebra is generated by
the time-translation and scaling symmetries
\[
[\X{trans},\X{scal}]=\X{trans} .
\]
In the conformal case of equations (NLW) and (mKdV-2,H)
this algebra is enlarged by an inversion symmetry
\[
[\X{trans},\X{inver}]=2\X{scal} ,\qquad
[\X{scal},\X{inver}]=\X{inver} .
\]
For the conformal (NLS) equation, its admitted algebra is
a central extension of the algebra for the conformal (NLW) equation,
as generated by the phase rotation symmetry
\[
[\X{phase},\X{trans}]=[\X{phase},\X{scal}]=[\X{phase},\X{inver}]=0 ,
\]
while the admitted algebra for the non-conformal (NLS) equation
as well as for the (dNLS) and \mbox{(dNLS-H)} equations is
a similar central extension of the algebra
for the non-conformal (NLW) equation.

\newpage

{\begin{center}\refstepcounter{tabul}\label{TablePintSymGroups}
Table~\thetabul. Point symmetry groups
\\[1.5ex] \scriptsize
\setcounter{tbn}{0}
\begin{tabular}{|l|l|l|}
\hline\vspacebefore
\hfill Generators $X\hfill$ & \hfill Group $\mathcal G\hfill$   &\hfill {Equation\hfill} \\
\hline\vspacebefore
$\X{trans}$, $\X{scal}$ &$U(1) \rtimes U(1)$ solvable
& \begin{tabular}{l}(mKdV-1), non-conformal (NLW),\\ non-conformal (mKdV-2,H)\end{tabular}  \\
\hline\vspacebefore
$\X{trans}$, $\X{scal}$, $\X{inver}$ & $SL(2,{\mathbb R})$ semisimple & \begin{tabular}{l}conformal (NLW),\\ conformal (mKdV-2,H)\end{tabular} \\
\hline\vspacebefore
$\X{trans}$, $\X{scal}$, $\X{phase}$ & \begin{tabular}{l}$(U(1) \rtimes U(1)) \times U(1)$\\ solvable\end{tabular}
& \begin{tabular}{l}non-conformal (NLS),\\ (dNLS), (dNLS-H)\end{tabular}\\
\hline\vspacebefore
$\X{trans}$, $\X{scal}$, $\X{inver}$, $\X{phase}$
& \begin{tabular}{l}$SL(2,{\mathbb R})\times U(1)$ centrally \\
extended semisimple \end{tabular} &
conformal (NLS)\\
\hline
\end{tabular}
\end{center}}

\section{Optimal symmetry groups for invariant solutions}

All one-dimensional point symmetry subgroups ${\mathcal G}_{(1)}$
determine corresponding group-invariant solutions $u(t,r)$
when the above wave equations
are augmented by the invariant surface condition $X u=0$
where $X$ is the symmetry generator of ${\mathcal G}_{(1)}$
in characteristic form.
In particular, the invariant surface condition reduces a wave equation~(WE)
to an ODE \cite{Olver1986,Bluman&Anco2002}.
Any two conjugate subgroups will give rise to reduced ODEs
that are related by a conjugacy transformation
in the full point symmetry group ${\mathcal G}$
acting on the invariant solutions $u(t,r)$ determined by each subgroup.
Hence, up to the action of ${\mathcal G}$,
all invariant solutions for a given wave equation
can be obtained by selecting a one-dimensional subgroup
in each conjugacy class of all admitted one-dimensional point symmetry
subgroups ${\mathcal G}_{(1)}$.
Such a selection is called an
{\em optimal set of subgroups}~\cite{Ovsiannikov1982}.

Optimal subalgebras have been classified in the work
presented in~\cite{Patera&Winternitz1977}
for all three- and four-dimensional Lie algebras.
Applying this classification to the point symmetry algebras of
the wave equations under consideration,
we have the following list of optimal one-dimensional subalgebras
for finding all invariant solutions of equations
(NLW), (NLS), (dNLS) and (dNLS-H), (mKdV-1,2,H) for $m>0$:

{\begin{center}\refstepcounter{tabul}\label{TableOptSubalgebras}
Table~\thetabul. Optimal one-dimensional subalgebras\\
($a,b,c$ are arbitrary constants)
\\[1.5ex] \scriptsize
\setcounter{tbn}{0}
\begin{tabular}{|l|l|}
\hline\vspacebefore
\hfill {Generators \hfill}   &\hfill {Equation\hfill} \\
\hline\vspacebeforemore
$\X{trans}$, $\X{scal}$ & \begin{tabular}{l}(mKdV-1), non-conformal (NLW),\\ non-conformal (mKdV-2,H)\end{tabular}\\
\hline\vspacebeforemore
$\X{trans}$, $\X{scal}$, $\X{inver}+\X{trans}$ & conformal (NLW), conformal (mKdV-2,H) \\
\hline\vspacebefore
$\X{phase}$, $\X{scal}+a\X{phase}$, $\X{trans}+b\X{scal}+c\X{phase}$ & non-conformal (NLS), (dNLS), (dNLS-H)\\
\hline\vspacebefore
\begin{tabular}{l}$\X{phase}$, $\X{trans}$, $\X{trans}\pm\X{scal}$,\\ $\X{scal}+a\X{phase}$, $\X{trans}+\X{inver}+b\X{phase}$\end{tabular}
& conformal (NLS)\\
\hline
\end{tabular}
\end{center}}

Invariant solutions for (NLW), (NLS), (dNLS)
with $m\geq 0$ (i.e. in all dimensions $n\geq1$)
have been derived in~\cite{Anco&Liu2004,Polyanin&Zaitsev2002,Ibragimov1994V1,Fushchych&Moskaliuk1981,Fushchych&Shtelen&Serov,Cherniga1995,Kaup&Newell1978}.
We will present invariant solutions of (dNLS-H), (mKdV-1,2,H)
for $m>0$ (i.e. in dimensions $n>1$) elsewhere.
(The $m=0$ case of (mKdV-1,2,H) is treated in~\cite{Olver1986,Miura1976,Gardner&Greene&Kruskal&Miura1974,Polyanin&Zaitsev2002}).

\section{ Local conservation laws }

As the wave equations~(NLW), (NLS) and (dNLS) each have a Lagrangian formulation,
all their admitted variational point symmetries yield corresponding
conserved quantities which are well known~\cite{Strauss1989}
for (NLW) and (NLS).
In particular,
time translation symmetry yields energy,
space translation and boost symmetries yield momenta,
and inversion symmetry yields conformal energy,
while phase rotation symmetry yields charge.
In addition, there is a special nonlinearity power for which
the scaling symmetry becomes variational and yields a dilational energy.
Our classification in the following tables
%~\ref{TableLocalCLsNLW},~\ref{TableLocalCLsNLS},~\ref{TableLocalCLsdNLS}
provides a completeness result
that no additional conserved quantities up to first order
are admitted by these equations~(NLW), (NLS) and (dNLS)
for special nonlinearity powers
(excluding all linear cases i.e. $p=0$; plus $p=1$ for (NLW)).

The remaining wave equations (dNLS-H) and (mKdV-1,2,H)
do not have a Lagrangian formulation,
so consequently their admitted conservation laws
come from multipliers given by adjoint-symmetries 
\cite{Anco&Bluman2002b}
rather than symmetries.
For the Hamiltonian equations (dNLS-H) and (mKdV-H),
note the Hamiltonian itself provides one conservation law.
Our results in the following tables
%~\ref{TableLocalCLsdNLS-H},~\ref{TableLocalCLsmKdV-1},~\ref{TableLocalCLsmKdV-2},~\ref{TableLocalCLsmKdV-H}
for all these radial equations with $m>0$ are new.

\setcounter{tbn}{0}
{\begin{center}\refstepcounter{tabul}\label{TableLocalCLsNLW}
Table~\thetabul. Local conservation laws for (NLW)
\\[1.5ex] \scriptsize
\setcounter{tbn}{0}
\begin{tabular}{|l|l|l|l|}
\hline\vspacebefore
\hfill $\Psi^t\hfill$ &\hfill $\Psi^r\hfill$  &\hfill {Remarks\hfill} \\
\hline{\raisebox{0ex}[3.8ex][0ex]{\null}} $\dfrac{1}2(u_t^2+u_r^2)\mp F(u,p+1)$ & $-u_tu_r$ & energy \\[0.8ex]
%1
\hline{\raisebox{0ex}[3.8ex][0ex]{\null}} $u_tu_r$ & $-\dfrac12(u_t^2+u_r^2)\mp F(u,p+1)$ & momentum, $m=0$\\[0.8ex] %radial
%2
\hline{\raisebox{0ex}[3.8ex][0ex]{\null}}
         $\dfrac12 r(u_t^2+u_r^2)+tu_tu_r\mp rF(u,p+1)$ & $\begin{array}{l}-\dfrac12 t(u_t^2+u_r^2)-ru_tu_r\\ \mp tF(u,p+1)\end{array}$
         & \begin{tabular}{l}boost momentum,\\ $m=0$\end{tabular}\\[0.8ex] %boost
%3
\hline{\raisebox{0ex}[6.1ex][0ex]{\null}}
                             $\renewcommand{\arraycolsep}{0ex}
                             \begin{array}{l}\dfrac12t(u_t^2+u_r^2)+ru_tu_r\\+\dfrac2{p-1}uu_t\mp tF(u,p+1)\end{array}$
                           & $\renewcommand{\arraycolsep}{0ex}
                             \begin{array}{l}-\dfrac12r(u_t^2+u_r^2)-tu_tu_r\\
                             -\dfrac{2}{p-1}uu_r\mp rF(u,p+1)\end{array}$ &\begin{tabular}{l} dilational energy,\\ $m=\dfrac4{p-1}$\end{tabular}\\[3.3ex]
%4
\hline{\raisebox{0ex}[6.3ex][0ex]{\null}}
                             $\renewcommand{\arraycolsep}{0ex}
                             \begin{array}{l} \dfrac12(t^2+r^2)(u_t^2+u_r^2)+2tru_tu_r-\dfrac{2}{p-1}u^2\\[1ex]
                             +\dfrac4{p-1}tuu_t\mp\left(t^2+r^2\right)F(u,p+1)\end{array}$&
                             $\begin{array}{l} -tr(u_t^2+u_r^2)-(t^2+r^2)u_tu_r\\[1ex]
                             -\dfrac{4}{p-1}tuu_r\mp2trF(u,p+1)\end{array}$& \begin{tabular}{l}conformal energy,\\ $m=\dfrac4{p-1}$\end{tabular}\\[1ex]
%5
\hline
\end{tabular}
\end{center}}

\newpage

\setcounter{tbn}{0}
{\begin{center}\refstepcounter{tabul}\label{TableLocalCLsNLS}
Table~\thetabul. Local conservation laws for (NLS)
\\[1.5ex] \scriptsize
\setcounter{tbn}{0}
\begin{tabular}{|l|l|l|}
\hline\vspacebefore
\hfill $\Psi^t\hfill$ &\hfill $\Psi^r\hfill$  &\hfill {Remarks\hfill} \\
\hline\vspacebefore $|u|^2$ & $\i(u_r\bar u-u\bar u_r)$ & charge \\[0.5ex]
\hline\vspacebeforemore  $\dfrac12|u_r|^2\mp F(|u|, p+2)$ & $-\dfrac12(u_r\bar u_t-u_t\bar u_r)$& energy \\[1.1ex]
\hline\vspacebefore $\i t(u_r\bar u-u\bar u_r)-r|u|^2$
                           & $\begin{array}{l}\i t(u\bar u_t-u_t\bar u)+\i r(u\bar u_r-u_r\bar u)\\
                             +2t|u_r|^2\pm4t F(|u|, p+2)\end{array}$
                           &  boost momentum, $m=0$ \\
\hline\vspacebefore $\dfrac\i2(u_r\bar u-u\bar u_r)$
                   & $\begin{array}{l}\dfrac\i2(u\bar u_t-u_t\bar u)\\+|u_r|^2\pm2 F(|u|, p+2)\end{array}$
                   &  momentum, $m=0$ \\
\hline{\raisebox{0ex}[8.5ex][0ex]{\null}} $\begin{array}{l}\dfrac12t|u_r|^2 +\dfrac{\i}8 r(u_r\bar u-u\bar u_r)\\\mp t F(|u|, p+2)
                           \end{array}$ & $\begin{array}{l}
                           -\dfrac12t(u_t\bar u_r+u_r\bar u_t)-\dfrac14r|u_r|^2\\[0.8ex]
                           +\dfrac{\i}8r(u_t\bar u-u\bar u_t)-\dfrac1{2p}(u\bar u_r+ u_r\bar u)\\[0.3ex] \mp\dfrac12 r F(|u|, p+2)
                           \end{array} $& dilational energy, $m=\dfrac4p -1$\\[5.8ex]
\hline{\raisebox{0ex}[8.5ex][0ex]{\null}}  $\begin{array}{l}
                           \dfrac12t^2|u_r|^2-\dfrac18r^2|u|^2\\
                            +\dfrac{\i}4 tr(u_r\bar u-u\bar u_r)\\
                           \mp t^2 F(|u|, p+2)
                           \end{array}$
                           & $\begin{array}{l}
                           -\dfrac12t^2(u_t\bar u_r+u_r\bar u_t)-\dfrac12t|u_r|^2\\[0.8ex]
                           +\dfrac{\i}8 r^2(u_r\bar u-u\bar u_r)+\dfrac{\i}4 tr(u_t\bar u-u\bar u_t)\\[0.8ex]
                           -\dfrac{1}{2p} t(u_r\bar u+u\bar u_r)\mp tr F(|u|, p+2)%\\[0.8ex]
                           \end{array} $& conformal energy, $m=\dfrac4p -1$\\
\hline
\end{tabular}
\end{center}}

%\newpage

\setcounter{tbn}{0}
{\begin{center}\refstepcounter{tabul}\label{TableLocalCLsdNLS}
Table~\thetabul. Local conservation laws for (dNLS)
\\[1.5ex] \scriptsize
\setcounter{tbn}{0}
\begin{tabular}{|l|l|l|}
\hline\vspacebefore
\hfill $\Psi^t\hfill$ &\hfill $\Psi^r\hfill$  &\hfill {Remarks\hfill} \\
\hline\vspacebefore $|u|^2$ & $\i u_r\bar u-\i u\bar u_r\mp 2F(|u|,p+2)$ & charge\\
\hline\vspacebefore $|u_r|^2\mp\i(u_r\bar u-u\bar u_r)|u|^{-2}F(|u|,p+2)$
                           & $\renewcommand{\arraycolsep}{0ex}\begin{array}{l}
                           -(u_r\bar u_t+u_t\bar u_r)\\ \mp\i(u_t\bar u-u\bar u_t)|u|^{-2}F(|u|,p+2)\end{array}$ & energy \\
\hline\vspacebeforemore $\dfrac\i2(u_r\bar u-u\bar u_r)$ & $\dfrac\i2(u\bar u_t-u_t\bar u)+|u_r|^2$ &  $m=0$  \\[0.8ex]
\hline {\raisebox{0ex}[5.9ex][0ex]{\null}}$\renewcommand{\arraycolsep}{0ex}\begin{array}{l}
                             \dfrac\i2(u_r\bar u-u\bar u_r)+2t|u_r|^2\\
                             \mp2\i t(u_r\bar u-u\bar u_r)|u|^{-2}F(|u|,p+2)\end{array}$ &
                             $\renewcommand{\arraycolsep}{0ex}\begin{array}{l}
                             \dfrac\i2(u_t\bar u-u\bar u_t)-2t(u_r\bar u_t+u_t\bar u_r)\\
                             \pm2\i t(u_t\bar u-u\bar u_t)|u|^{-2}F(|u|,p+2)\\ -\dfrac1p(u_r\bar u+u\bar u_r)-r|u_r|^2
                             \end{array}$ &  \begin{tabular}{l} dilational energy,\\ $m=\dfrac2p -1$\end{tabular}\\[3.4ex]
\hline
\end{tabular}
\end{center}}

\setcounter{tbn}{0}
{\begin{center}\refstepcounter{tabul}\label{TableLocalCLsdNLS-H}
Table~\thetabul. Local conservation laws for (dNLS-H)
\\[1.5ex] \scriptsize
\setcounter{tbn}{0}
\begin{tabular}{|l|l|l|}
\hline\vspacebefore
\hfill $\Psi^t\hfill$ &\hfill $\Psi^r\hfill$  &\hfill {Remarks\hfill} \\
\hline\vspacebeforemore $r^{-m/2}(u+\bar u)$
 & $r^{-m/2}(\i(\bar u_r -u_r)\mp|u|^{p}(u+\bar u)+\dfrac{m\i}{2}r^{-1}(\bar u-u))$ & \\[0.5ex]
\hline\vspacebeforemore $r^{-m/2}\i (\bar u-u)$ & $r^{-m/2}(u_{r}+\bar u_r \pm\i |u|^{p}(u-\bar u)+\dfrac{m}{2}r^{-1}(u+\bar u))$ & \\[0.8ex]
\hline{\raisebox{0ex}[5ex][0ex]{\null}} $\renewcommand{\arraycolsep}{0ex}\begin{array}{l}
                           \dfrac\i2(u\bar u_r-u_r\bar u)\\
                           \pm2F(|u|,p+2)
                           \end{array}$ &
                           $\renewcommand{\arraycolsep}{0ex}\begin{array}{l}
                           \dfrac\i2(u_t\bar u-u\bar u_t)-|u_r|^2 -\dfrac{m^2}{4} r^{-2}|u|^2\\
                            -|u|^{p+2}\mp\i|u|^p(u\bar u_r -u_r\bar u)\end{array}$ & Hamiltonian \\[0.5ex]
\hline\vspacebefore $|u|^2$ & $\i(u\bar u_r-u_r\bar u)\mp2(p+1) F(|u|,p+2)$ & charge, $m=0$ \\[0.5ex]
\hline{\raisebox{0ex}[8ex][0ex]{\null}} $\renewcommand{\arraycolsep}{0ex}\begin{array}{l}
                           \dfrac{\i}2 t(u\bar u_r-u_r\bar u)\\
                           \pm2 tF(|u|,p+2)+\dfrac r2|u|^2
                           \end{array}$ &
                           $\renewcommand{\arraycolsep}{0ex}\begin{array}{l}
                           \dfrac \i 2t(u_t\bar u-u\bar u_t)-t(|u_r|^2+\dfrac{m^2}4 r^{-2}|u|^2)\\[1ex]
                           -t(|u|^{p+2}\pm\i|u|^p(u\bar u_r -u_r\bar u))\\[1ex]
                           +\dfrac{\i}2 r(u_r\bar u-u\bar u_r)\mp (p+1)rF(|u|,p+2)\end{array} $ & $m=\dfrac2p -2$ \\[5.8ex]
\hline
\end{tabular}
\end{center}}

\newpage

\setcounter{tbn}{0}
{\begin{center}\refstepcounter{tabul}\label{TableLocalCLsmKdV-H}
Table~\thetabul. Local conservation laws for (mKdV-H), $m\ne0$
\\[1.5ex] \scriptsize
\setcounter{tbn}{0}
\begin{tabular}{|l|l|l|}
\hline\vspacebefore
\hfill $\Psi^t\hfill$ &\hfill $\Psi^r\hfill$  &\hfill {Remarks\hfill} \\
\hline\vspacebeforemore
 $r^{-m/2}u$ & $-r^{-m/2}\left(u_{rr}+mr^{-1}u_r\pm u^{p+1}\right)$ & mass  \\[0.8ex]
\hline{\raisebox{0ex}[6ex][0ex]{\null}} $\dfrac12u_r^2\mp F(u,p+2)$ &
                           $\renewcommand{\arraycolsep}{0ex}\begin{array}{l}
                           -u_tu_r+\dfrac12u_{rr}^2+mr^{-1}u_ru_{rr}\\
                           \pm u^{p+1}u_{rr}+\dfrac{m^2}{2} r^{-2}u_r^2
                           \pm m r^{-1}u^{p+1}u_r+\dfrac12u^{2p+2}
                           \end{array}$ & Hamiltonian  \\[3.5ex]
\hline{\raisebox{0ex}[8.5ex][0ex]{\null}}
                           $\renewcommand{\arraycolsep}{0ex}\begin{array}{l}
                           \dfrac32tu_r^2-\dfrac12ru^2\mp3F(u,p+2)
                           \end{array}$ & $\renewcommand{\arraycolsep}{0ex}\begin{array}{l}
                           -3tu_tu_r+\dfrac32tu_{rr}^2+\dfrac32tu^{2p+2}\pm3 tu^{p+1}u_{rr}+ruu_{rr}\\[1ex]
                           -6\dfrac{p-2}{p} tr^{-1}u_ru_{rr}+6\dfrac{(p-2)^2}{p^2}t r^{-2}u_r^2-\dfrac12ru_r^2\\[1ex]
                           -2\dfrac{p+1}{p}uu_r\pm (p+1)rF(u,p+2)
                           \end{array}$ &  $m=\dfrac4p -2$\\[6.3ex]
\hline{\raisebox{0ex}[9.3ex][0ex]{\null}} $\renewcommand{\arraycolsep}{0ex}\begin{array}{l}
                           \dfrac92t^2u_r^2\mp3 t^2u^3-3tru^2\mp\\\mp6 tr^{-1}u\mp r^2u
                           \end{array} $ &
                           $\renewcommand{\arraycolsep}{0ex}\begin{array}{l}
                           -9t^2r^{-1}u_r u_t+\dfrac92t^2r^{-1}u_{rr}^2+18t^2r^{-2}u_ru_{rr}\pm3r^{-1} u\\[1ex]
                           \pm9 t^2r^{-1}u^2u_{rr}+6tuu_{rr}\pm6 tr^{-2}u_{rr}+6tr^{-2}u^2\\[0.5ex]
                           +18t^2r^{-3}u_r^2-3tu_r^2\pm18t^2r^{-2}u^2u_r+r^2u^2\\[0.5ex]
                           \pm12 tr^{-2}u_r\mp u_r+\dfrac92t^2r^{-1}u^4\pm 4tu^3\pm r u_{rr}
                           \end{array} $  &  $m=2$, $p=1$ \\[7ex]
\hline
\end{tabular}\\
\end{center}}

\setcounter{tbn}{0}
{\begin{center}\refstepcounter{tabul}\label{TableLocalCLsmKdV-1}
Table~\thetabul. Local conservation laws for (mKdV-1), $m\ne0$
\\[1.5ex] \scriptsize
\setcounter{tbn}{0}
\begin{tabular}{|l|l|l|}
\hline\vspacebefore
\hfill $\Psi^t\hfill$ &\hfill $\Psi^r\hfill$  &\hfill {Remarks\hfill} \\
\hline\vspacebefore $r^{-m}u$  & $-r^{-m}(u_{rr}+mr^{-1}u_r\pm u^{p+1})$ &  mass \\[0.5ex]
\hline
\end{tabular}
\end{center}}

%\newpage

\setcounter{tbn}{0}
{\begin{center}\refstepcounter{tabul}\label{TableLocalCLsmKdV-2}
Table~\thetabul. Local conservation laws for (mKdV-2), $m\ne0$
\\[1.5ex] \scriptsize
\setcounter{tbn}{0}
\begin{tabular}{|l|l|l|}
\hline\vspacebefore
\hfill $\Psi^t\hfill$ &\hfill $\Psi^r\hfill$  &\hfill {Remarks\hfill} \\
\hline\vspacebefore $r^{-m} u$ & $-r^{-m}(u_{rr}+mr^{-1}u_r\pm u^{p+1})$ & mass  \\[0.5ex]
\hline\vspacebeforemore $\dfrac12r^{1/2}u^2$ & $-r^{1/2}uu_{rr}+\dfrac12r^{1/2}u_r^2-r^{-1/2}uu_r\mp\dfrac{3}{4} r^{1/2}u^{4}
\pm\dfrac{1}8r^{-3/2}u^2$ &\begin{tabular}{l} dilational momentum,\\ $m=3/2$, $p=2$\end{tabular}\\[0.8ex]
\hline\vspacebeforemore $\dfrac12ru^2$ & $-ruu_{rr}+\dfrac12ru_r^2-2uu_r\mp\dfrac{3}{4} ru^{4}$ & \begin{tabular}{l}
   dilational momentum,\\ $m=3$, $p=2$\end{tabular}\\[0.8ex]
\hline
\end{tabular}
\end{center}}

For completeness we mention that in the $m=0$ case 
the well known mKdV conservation laws 
are listed 
in~\cite{Miura&Gardner&Kruskal1968,Miura1976,Olver1986,Anco&Bluman2002a}.

%\setcounter{tbn}{0}
%{\begin{center}\refstepcounter{tabul}\label{TableLocalCLsmKdV}
%Table~\thetabul. Local conservation laws for (mKdV-1,\,2,\,H), $m=0$
%with second-order density
%\\[1.5ex] \scriptsize
%\setcounter{tbn}{0}
%\renewcommand{\arraystretch}{1.2}
%\begin{tabular}{|l|l|l|}
%\hline\vspacebefore
%\hfill $\Psi^t\hfill$ &\hfill $\Psi^r\hfill$  &\hfill {Remarks\hfill} \\
%\hline\vspacebefore $u$ & $-u_{rr}\mp u^{p+1}$ & mass  \\
%\hline{\raisebox{0ex}[3.9ex][0ex]{\null}} $\dfrac12 u^2$ & $uu_{rr}-\dfrac12u_r^2\pm(p+1)F(u,p+2)$ & momentum\\[0.8ex]
%\hline{\raisebox{0ex}[3.9ex][0ex]{\null}} $\dfrac12u_r^2\mp F(u,p+2)$ &
%                    $-u_tu_r+\dfrac12u_{rr}^2\pm u^{p+1}u_{rr}+\dfrac{m^2}{2} r^{-2}u_r^2+\dfrac12u^{2p+2}$ & energy  \\[0.8ex]
%\hline\vspacebeforemore $tu^2\pm ru$ &
%                           $u_{rr}(-2tu \mp r)+ tu_r^2\pm u_r\mp \dfrac43tu^3- ru^2$ & Galilean momentum,  $p=1$ \\[0.8ex]
%\hline
%\end{tabular}
%\end{center}}

\section{ Norms and critical powers }

The wave equations (NLW), (NLS), (dNLS) and (dNLS-H), (mKdV-H)
possess both a scaling symmetry (which is uniform in $m$)
and a conserved energy or Hamiltonian,
so they each have an associated critical power
with respect to the energy norm,
$E[u]=\int_0^\infty e[u] r^m dr$,
as shown in the following table.

\newpage

\setcounter{tbn}{0}
{\begin{center}\refstepcounter{tabul}\label{TableEnergyNorms}
Table~\thetabul. Energy norms %$m\ne0$
\\[1.5ex] \scriptsize
\setcounter{tbn}{0}
\begin{tabular}{|l|l|l|}
\hline\vspacebefore
\hfill {$E[u]$\hfill} &\hfill critical power $p$   &\hfill {Remarks\hfill} \\
\hline\vspacebeforemore
{$\int_0^\infty$}$ \Bigl(\dfrac{1}{2}(|u_r|^2 +|u_t|^2) \mp F(u,p+1)\Bigr) r^m dr$
& $1+\dfrac{4}{m-1}$ & (NLW), $m\neq1$ \\[0.8ex]
\hline\vspacebeforemore
{ $\int_0^\infty$}$ \Bigl(\dfrac{1}{2} |u_r|^2 \mp F(|u|,p+1)\Bigr) r^m dr$
& $\dfrac{4}{m-1}$ & (NLS), $m\neq1$ \\[0.8ex]
\hline\vspacebeforemore
{ $\int_0^\infty$}$ \Bigl(\dfrac{1}{2} |u_r|^2 \mp\i(u\bar u_r -\bar u u_r)|u|^{-2}F(|u|,p+2)\Bigr) r^m dr$
& $\dfrac{2}{m-1}$& (dNLS), $m\neq1$ \\[0.8ex]
\hline\vspacebeforemore
{ $\int_0^\infty$}$\Bigl(\dfrac{\i}{2}(u\bar u_r -\bar u u_r)\pm2F(|u|,p+2)\Bigr) r^m dr$
& $\dfrac{2}{m}$ & (dNLS-H), $m\neq0$ \\[0.8ex]
\hline\vspacebeforemore
{ $\int_0^\infty$}$ \Bigl(\dfrac{1}{2} |u_r|^2 \pm F(u,p+2)\Bigr) r^m dr$
& $\dfrac{4}{m-1}$ & (mKdV-H), $m\neq1$ \\[0.8ex]
\hline
\end{tabular}
\end{center}}

All these wave equations also possess dilational energies or dilational Hamiltonians
for special nonlinearity powers $p$ depending on $m$,
in addition to the well known conformal energies admitted for
the (NLW) and (NLS) equations in the case of conformal powers $p$
(cf. table~\ref{TableConfPowers}).
Interestingly, the Hamiltonian mKdV equation (mKdV-H) admits
a conformal energy in this case too.

\setcounter{tbn}{0}
{\begin{center}\refstepcounter{tabul}\label{TableDilationEnergy}
Table~\thetabul. Dilational energies and dilation powers
\\[1.5ex] \scriptsize
\setcounter{tbn}{0}
\begin{tabular}{|l|l|l|}
\hline\vspacebefore
\hfill {Dilational energy\hfill} &\hfill Dilation power $p$   &\hfill {Remarks\hfill} \\
\hline\vspacebeforemore
{ $\int_0^\infty$}$ \Bigl(t e[u] +\bigl(ru_r + \dfrac{m}{2} u\bigr)u_t\Bigr) r^m dr$ & $1+\dfrac{4}{m}$ & (NLW), $m\neq0$ \\[0.8ex]
\hline\vspacebeforemore
{ $\int_0^\infty$}$ \Bigl(t e[u] +\dfrac{\i}{4} r(\bar u u_r -\bar u_r u) \Bigr)r^m dr$
& $\dfrac{4}{m+1}$ & (NLS), $m\neq-1$ \\[0.8ex]
\hline\vspacebeforemore
{ $\int_0^\infty$}$ \Bigl(t e[u] +\dfrac{\i}{4} r(\bar u u_r -\bar u_r u)\Bigr) r^m dr$
& $\dfrac{2}{m+1}$& (dNLS), $m\neq-1$ \\[0.8ex]
\hline\vspacebeforemore
{ $\int_0^\infty$}$ \Bigl(t e[u] +\dfrac{1}{4} r |u|^2\Bigr) r^m dr$
& $\dfrac{2}{m+2}$ & (dNLS-H), $m\neq-2$ \\[0.8ex]
\hline\vspacebeforemore
{ $\int_0^\infty$}$ \Bigl(t e[u] -\dfrac{1}{6} r u^2\Bigr) r^m dr$
& $\dfrac{4}{m+2}$ & (mKdV-H), $m\neq-2$ \\[0.8ex]
\hline
\end{tabular}
\end{center}}

%\newpage

\setcounter{tbn}{0}
{\begin{center}\refstepcounter{tabul}\label{TableConformalEnergy}
Table~\thetabul. Conformal energies and conformal powers
\\[1.5ex] \scriptsize
\setcounter{tbn}{0}
\begin{tabular}{|l|l|l|}
\hline\vspacebefore
\hfill {Conformal energy\hfill} &\hfill Conformal power $p$   &\hfill {Remarks\hfill} \\
\hline\vspacebeforemore
{ $\int_0^\infty$}$ \Bigl((t^2+r^2) e[u] +2t\bigl(ru_r + \dfrac{m}{2} u\bigr)u_t\Bigr) r^m dr$ & $1+\dfrac{4}{m}$ & (NLW), $m\neq0$ \\[0.8ex]
\hline\vspacebeforemore
{ $\int_0^\infty$}$ \Bigl(t^2 e[u] +\dfrac{\i}{2} tr(\bar u u_r -\bar u_r u) -\dfrac{1}{4} r^2 |u|^2\Bigr) r^m dr$
& $\dfrac{4}{m+1}$ & (NLS), $m\neq-1$ \\[0.8ex]
\hline\vspacebeforemore
{ $\int_0^\infty$}$ \Bigl(t^2 e[u] -\dfrac{1}{3} \bigl(tr u^2 +\dfrac{1}{2} tr^{-1} u +\dfrac{1}{3} r^2 u\bigr)\Bigr) r^2 dr$
& $1$ & (mKdV-H), $m=2$ \\[0.8ex]
\hline
\end{tabular}
\end{center}}

An interesting pattern in tables~18 and~19 is that 
when $m$ is expressed in terms of $p$
then the difference $m_{\rm crit.} -m_{\rm dil.}$ (for any fixed $p$) is equal to
${\rm ord}(\partial_r) - {\rm ord}(\partial_t) +1 >0$
where ``${\rm ord}$'' refers to the highest order of a specified derivative
appearing in the wave equation.
Accordingly, the dilation and conformal powers are subcritical in all cases.

Two other norms of analytical interest are the radial $L^2$ norm
and the radial $H^s$ norm given by
$||u||_{L^2} =\left( \int_0^\infty |u|^2 r^m dr \right)^{1/2}$
and
$||u||_{H^s} =\left( \int_0^\infty |\partial^s_r u|^2 r^m dr \right)^{1/2}$
for any positive integer $s$.
The latter norm has a natural extension to all $s\geq 0$
defined in terms of the Fourier transform
$\hat u = \int_{\mathbb{R}^{m+1}} u(t,|x|) \exp(-k\cdot x) d^{m+1}x$
such that $u$ is in $H^s$ iff $(1+|k|^s)\hat u$ is in $L^2(\mathbb{R}^{m+1})$.
The following tables list the critical powers $p$
for which these norms are scaling-invariant.
Note in the case of the Schr\"odinger equations (NLS), (dNLS) and (dNLS-H),
the $L^2$ norm coincides with the conserved charge.

\setcounter{tbn}{0}
{\begin{center}\refstepcounter{tabul}\label{TableL2CritPower}
Table~\thetabul. $L^2$ critical powers %$m\ne0$
\\[1.5ex] \scriptsize
\setcounter{tbn}{0}
\begin{tabular}{|l|l|}
\hline\vspacebefore
\hfill critical power $p$   &\hfill {Remarks\hfill} \\
\hline\vspacebeforemore
$1+\dfrac{4}{m+1}$ & (NLW)\\[0.8ex]
\hline\vspacebeforemore
$\dfrac{4}{m+1}$ & (NLS), (mKdV-1,2,H)\\[0.8ex]
\hline\vspacebeforemore
$\dfrac{2}{m+1}$ & (dNLS), (dNLS-H)\\[0.8ex]
\hline
\end{tabular}
\end{center}}

%\newpage

\setcounter{tbn}{0}
{\begin{center}\refstepcounter{tabul}\label{TableHsCritPower}
Table~\thetabul. $H^s$ critical powers %$m\ne0$
\\[1.5ex] \scriptsize
\setcounter{tbn}{0}
\begin{tabular}{|l|l|l|}
\hline\vspacebefore
\hfill  critical power $p$ &\hfill critical $s\hfill$ &\hfill {Remarks\hfill} \\
\hline\vspacebeforemore
$1+\dfrac{4}{m+1-2s}$ & $\dfrac{m+1}{2} -\dfrac{2}{p-1}$ & (NLW)\\[1ex]
\hline\vspacebeforemore
$\dfrac{4}{m+1-2s}$ & $\dfrac{m+1}{2} -\dfrac{2}{p}$ & (NLS), (mKdV-1,2,H)\\[1ex]
\hline\vspacebeforemore
$\dfrac{2}{m+1-2s}$ & $\dfrac{m+1}{2} -\dfrac{1}{p}$ & (dNLS), (dNLS-H)\\[1ex]
\hline
\end{tabular}
\end{center}}

\section{Concluding remarks}

The utility of symmetries and conservation laws can be extended by
means of potential systems
\cite{Bluman&Kumei1989,Bluman&Cheviakov,Bluman&etal2006}.
A {\em potential system} for a radial wave equation~(WE) arises from
any conservation law such that vanishing set of its multiplier,
$Q=0$,
is contained in the set of all formal solutions~$u$ of the given equation.
Potentiating such a conservation law yields the system
\[
v_t=r^m \Psi^r ,\qquad v_r=-r^m \Psi^t
\]
whose solutions $v$ up to shifts 
($v \rightarrow v+c$ for an arbitrary constant $c$)
are in one-to-one correspondence with
the set of solutions $u$.
For a given potential system,
any admitted symmetry or conservation law that has an essential dependence
on the potential $v$
represents a {\em nonlocal symmetry} or {\em nonlocal conservation law},
respectively,
of the wave equation~(WE).
Of course,
the superposition of a local symmetry or a local conservation law 
with a nonlocal one 
yields further nonlocal ones,
and so for the purpose of classifications
we will mod out the admitted sets of local symmetries and conservation laws.

All potential systems arising from the conservation laws 
for the wave equations 
in tables~\ref{TableLocalCLsNLW} to~\ref{TableLocalCLsmKdV-2}
are given by potentiating:
the mKdV equations themselves (mKdV-1,2,H)
and the Hamiltonian variant of the derivative Schr\"odinger equation itself (dNLS-H);
the charge conservation law for the Schr\"odinger equations
(NLS), (dNLS), and (dNLS-H) in the case $m=0$;
%the momentum conservation law for the $m=0$ mKdV equations (mKdV-1,2,H);
and the dilational momentum conservation laws for the mKdV equation (mKdV-2)
in the cases $m=3,\frac{3}{2}$.
We find that only the first of these potential systems
--- potentiation of the (mKdV-1) equation itself 
(including the case $m=0$) ---
yields nonlocal conservation laws
and none yield any nonlocal symmetries.

%\newpage

\setcounter{tbn}{0}
{\begin{center}\refstepcounter{tabul}\label{TablePotCLsmKdV-1}
Table~\thetabul. Nonlocal conservation laws for (mKdV-1)\\
$\varepsilon =\sqrt{\mp 1}=\i,1$ respectively
in the defocusing/focusing cases
\\[1.5ex] \scriptsize
\setcounter{tbn}{0}
\begin{tabular}{|l|l|l|l|}
\hline\vspacebefore
\hfill {Potential system\hfill} &\hfill $\Psi^t\hfill$ &\hfill $\Psi^r\hfill$  &\hfill {Remarks\hfill} \\
\hline\vspacebefore $\begin{array}{l}v_r=u,\\ v_t=u_{rr}-\varepsilon^2u^{3}\end{array}$ & $e^{\varepsilon\sqrt2v}$
  & $\dfrac13e^{\varepsilon\sqrt2v}(u^2-\sqrt2u_r)$ & $m=0$,  $p=2$  \\
\hline\vspacebefore $\begin{array}{l}v_r=u,\\ v_t=u_{rr}+\dfrac32r^{-1}u_r-\varepsilon^2u^{3}\end{array}$ & $re^{\varepsilon\sqrt2v}$
  & $\dfrac13e^{\varepsilon\sqrt2v}\left(ru^2-\dfrac{\sqrt2}2(2ru_r+u)\right)$ & $m=3/2$,  $p=2$  \\[2.5ex]
\hline\vspacebefore $\begin{array}{l}v_r=u,\\ v_t=u_{rr}+3r^{-1}u_r-\varepsilon^2u^{3}\end{array}$ & $r^2e^{\varepsilon\sqrt2v}$
  & $\dfrac13e^{\varepsilon\sqrt2v}(r^2u^2-\sqrt2(r^2u_r+ru)+1)$ & $m=3/2$,  $p=2$  \\
\hline
\end{tabular}
\end{center}}

Each of these nonlocal conservation laws gives rise to a further potential system.
Of the three, none yield any additional nonlocal conservation laws,
and only the one case $m=0$
yields a~nonlocal symmetry (previously found in \cite{Galas1992,Guthrie&Hickman1993}).

%\setcounter{tbn}{0}
%{\begin{center}\refstepcounter{tabul}\label{TablePotSymKdV}
%Table~\thetabul. Nonlocal symmetries for (mKdV) $m=0$\\
%$\varepsilon =\sqrt{\mp 1}=\i,1$ respectively
%in the defocusing/focusing cases
%\\[1.5ex] \scriptsize
%\setcounter{tbn}{0}
%%\renewcommand{\arraystretch}{1.1}
%\begin{tabular}{|l|l|}
%\hline\vspacebefore
%\hfill {Potential system\hfill} & \hfill {Generator $X$\hfill} \\
%\hline{\raisebox{0ex}[3.9ex][0ex]{\null}}
%$v_r=u$, $w_r=e^{\varepsilon\sqrt2v}$, $w_t=\dfrac13e^{\varepsilon\sqrt2v}(\sqrt2u_r-u^2)$
%& $e^{\varepsilon\sqrt2v} \partial_u+\varepsilon w\partial_v +\dfrac{\sqrt2}6w^2\partial_w$\\[0.8ex]
%\hline
%\end{tabular}\end{center}}

Nevertheless, these potential systems may be very useful for 
finding new exact solutions \cite{Anco&Liu2004} 
of the equations (NLS), (dNLS), (dNLS-H), (mKdV-1,2,H), 
which we will pursue elsewhere.

\subsection*{Acknowledgments}
SCA is supported by an NSERC grant.
NMI acknowledges financial support from NSERC and
the Department of Mathematics of the University of British Columbia.
This research was partially supported by the grant of the President of Ukraine for young scientists (project number GP/F11/0061Â).

The authors are indebted to Thomas Wolf for extending the capabilities of
his computer program {\sc Crack} to handle the type of overdetermined systems
arising in this research, in particular where unknown parameters appear
not only polynomially but also in exponents.
Lauren Kowtecky is thanked for assistance with the computations
using {\sc Crack}.


\begin{thebibliography}{99}

\bibitem{Anco&Bluman1997}
Anco~S.C. and Bluman~G.,
Direct construction of conservation laws from field equations,
{\it Phys. Rev. Lett.} 1997, V.78, 2869--2873.

\bibitem{Anco&Bluman2002a}
Anco~S.C. and Bluman~G.,
Direct construction method for conservation laws of partial differential equations.~I.
Examples of conservation law classifications,
{\it Euro. J. Appl. Math.}, 2002, V.13, 545--566.
%(math-ph/0108023)

\bibitem{Anco&Bluman2002b}
Anco~S.C. and Bluman~G.,
Direct construction method for conservation laws of partial differential equations.~II. General treatment,
{\it Euro. J. Appl. Math.}, 2002, V.13, 567--585.
%(math-ph/0108024)

\bibitem{Anco&Liu2004}
Anco S.C. and Liu S.,
Exact solutions of semilinear radial wave equations in n dimensions,
{\it J. Math. Analysis Appl.}, 2004, V.297, 317--342.

%\bibitem{Bluman1993}
%Bluman~G.,
%Use and construction of potential symmetries,
%{\it Math. Comput. Modelling}, V.18 (10), 1993, 1--14.

\bibitem{Bluman&Anco2002}
Bluman~G. and Anco~S.C.,
{\it Symmetry and Integration Methods for Differential Equations},
Springer Applied Mathematics Series V.154, 2002.

\bibitem{Bluman&Cheviakov}
Bluman G. and Cheviakov A., 
Framework for potential systems and nonlocal symmetries: Algorithmic approach, 
{\it J. Math. Phys.}, 2005, V.46, 123506 (19 pages).

\bibitem{Bluman&etal2006}
Bluman G., Cheviakov A., and Ivanova N., 
Framework for nonlocally related PDE systems and nonlocal symmetries: 
extension, simplification, and examples.  
To appear in J. Math. Phys. (2006).

%\bibitem{Bluman&Kumei1987}
%Bluman G. and Kumei S.,
%On invariance properties of the wave equation,
%{\it J. Math. Phys.}, 1987, V.28, 307--318.

\bibitem{Bluman&Kumei1989}
Bluman~G.W., Kumei~S., {\it Symmetries and Differential Equations}, Springer, New York, 1989.

%\bibitem{Bluman&Reid&Kumei1988}
%Bluman~G.W., Reid~G.J., and Kumei~S., 
%New classes of symmetries for partial differential equations,
%{\it J. Math. Phys.}, 1988, V.29, 806--811.

\bibitem{Bluman&etal}
Bluman~G.W., Temuerchaolu, and Anco~S.,
New conservation laws obtained directly from symmetry action on known
conservation laws,
{\it J. Math. Anal. Appl.}, 2006, V.322, 233--250.


%\bibitem{Calogero&Degasperis1982}
%Calogero F. and Degasperis, A.,
%{\it Spectral transform and solitons: tools to solve and investigate
%nonlinear evolution equations}, North-Holland Publishing Company, Amsterdam, 1982.

\bibitem{Cherniga1995}
Cherniga R.M.,
Symmetry properties and exact solutions of the nonlinear Schr\"odinger equation with a power nonlinearity containing a derivative,
{\it Ukrain. Fiz. Zh.}, 1995, V.40, 376--384.

\bibitem{Fushchych&Moskaliuk1981}
Fushchych~W.I. and Moskaliuk~S.S.,
On some exact solutions of the nonlinear Schr\"odinger equations in three spatial dimensions,
{\it Lett. Nuovo Cim.}, 1981, V.31, 571--576.

\bibitem{Fushchych&Shtelen&Serov}
Fushchych W.I., Shtelen W.M. and Serov N.I.,
{\it Symmetry Analysis and Exact Solutions of Equations of Nonlinear Mathematical Physics}, Kluwer, Dordrecht, 1993.

\bibitem{Galas1992}
Galas F.,
New non-local symmetries with pseudopotentials,
{\it J. Phys. A: Math. Gen.}, 1992, V.25, L981--L986.

\bibitem{Gardner&Greene&Kruskal&Miura1974}
Gardner C.S., Greene J.M., Kruskal M.D. and Miura R.M.,
Korteweg--de Vries equation and generalizations. VI. Methods for exact solution,
{\it Comm. Pure Appl. Math},  1974, V.27, 97--133.

\bibitem{Guthrie&Hickman1993}
Guthrie G.A. and Hickman M.S.
Nonlocal symmetries of the KdV equation,
{\it  J. Math. Phys.}, 1993, V.34, 193--205.

\bibitem{Ibragimov1994V1}
Ibragimov~N.H. (Editor),
{\it CRC Handbook on Lie Group Analysis of Differential Equations -- Symmetries, Exact Solutions and Conservation Laws},
V.1, Chemical Rubber Company, Boca Raton, FL, 1994.

\bibitem{Kaup&Newell1978}
Kaup D.J. and Newell A.C.,
An exact solution for a derivative nonlinear Schrodinger equation,
{\it J. Math. Phys.}, 1978, V.19, 798--801.

\bibitem{Miura1976}
Miura R.M.,
The Korteweg--de Vries equation: a survey of results,
{\it SIAM Rev.}, 1976, V.18, 412--459.

\bibitem{Miura&Gardner&Kruskal1968}
Miura R.M., Gardner C.S. and Kruskal M.D.,
Korteweg--de Vries equation and generalizations. II. Existence of conservation laws and constants of motion,
{\it J. Math. Phys.}, 1968, V.9, 1204--1209.

\bibitem{Olver1986}
Olver P., {\it Applications of Lie Groups to Differential Equations},
Springer-Verlag, New York, 1986.

\bibitem{Ovsiannikov1982}
Ovsiannikov,~L.V.,
{\it Group Analysis of Differential Equations},
Academic Press, New York, 1982.

\bibitem{Patera&Winternitz1977}
Patera~J. and Winternitz~P. 1977
Subalgebras of real three- and four-dimensional Lie algebras
{\it J. Math. Phys.}, V.18, 1449--1455.

\bibitem{Polyanin&Zaitsev2002}
Polyanin~A.D. and Zaitsev~V.F.,
{\it Handbook of Nonlinear Equations Of Mathematical Physics},
Moscow, Fizmatlit, 2002.

%\bibitem{Popovych&Ivanova2005ConsLaws}
%Popovych R.O. and Ivanova N.M.,
%Hierarchy of conservation laws of diffusion--convection equations,
%{\it J. Math. Phys.}, 2005, V.46, 043502 (math-ph/0407008).

%\bibitem{Rollins&Shivamoggi1994}
%Rollins D.K. and Shivamoggi B.K.
%Painleve property and group symmetries of the generalized Korteweg--de Vries equation,
%{\it Phys. Scripta}, 1994, V.49 261--263.

\bibitem{Strauss1989}
Strauss W.A.,
{\it Nonlinear wave equations}, CBMS V.73, AMS, 1989.

%\bibitem{Wahlquist&Estabrook1975}
%Wahlquist~H.D. and Estabrook~F.B.,
%Prolongation structures of nonlinear evolution equations,
%{\it J. Math. Phys.}, 1975, V.16, N~1, 1--7.


\bibitem{Wolf1993}
Wolf T.,
An efficiency improved program LIEPDE for determining Lie-symmetries of PDEs,
{\it Proceedings of ``Modern Group Analysis: advanced analytical
and computational methods in mathematical physics''} (Catania, Italy, October 1992), Kluwer Academic Publishers, 377--385, 1993.

\bibitem{Wolf2002b}
Wolf~T.,
A comparison of four approaches to the calculation of conservation laws,
{\it Euro. J. Appl. Math.}, 2002, V.13, Part 5, 129--152.

\bibitem{Wolf2002a}
Wolf~T.,
Crack, LiePDE, ApplySym and ConLaw, section 4.3.5 in:
Grabmeier J., Kaltofen E.\ and Weispfenning~V.\ (Eds.): {\it Computer Algebra Handbook}, Springer, 2002, 465--468.



\end{thebibliography}
\end{document}